\documentclass[preprint, amsmath]{revtex4}
\usepackage{graphicx}
\usepackage{bm}

\newcommand{\ket}[1]{\ensuremath{|{#1}\rangle}}
\newcommand{\bra}[1]{\ensuremath{\langle{#1}|}}
\newcommand{\Do}{D$^0$}
\newcommand{\Dox}{D$^0$X}
\newcommand{\Aox}{A$^0$X}
\newcommand{\PI}{\ensuremath{\pi}}

\begin{document}
\title{Ultrafast control of donor-bound electron spins with single detuned optical pulses}
\par
\author{Kai-Mei C. Fu}
\affiliation{Information and Quantum Systems Laboratory, Hewlett-Packard
Laboratories, 1501 Page Mill Road, MS1123, Palo Alto, California
94304}
\author{Susan M. Clark}
\affiliation{Edward L. Ginzton Laboratory, Stanford University,
Stanford, California 94305-4088, USA}
\author{Charles Santori}
\affiliation{Information and Quantum Systems Laboratory, Hewlett-Packard
Laboratories, 1501 Page Mill Road, MS1123, Palo Alto, California
94304}
\author{Colin R. Stanley}
\affiliation{Department of Electronics and Electrical Engineering, Oakfield Avenue, University of Glasgow, Glasgow, G12 8LT, United Kingdom}
\author{M. C. Holland}
\affiliation{Department of Electronics and Electrical Engineering, Oakfield Avenue, University of Glasgow, Glasgow, G12 8LT, United Kingdom}
\author{Yoshihisa Yamamoto}
\affiliation{Edward L. Ginzton
Laboratory, Stanford University, Stanford, California 94305-4088,
USA}
\affiliation{National Institute of Informatics, 2-1-2 Hitotsubashi,
Chiyoda-ku, Tokyo 101-8430, Japan}
\maketitle

The ability to control spins in semiconductors is important in a variety of fields including spintronics and quantum information processing.  Due to the potentially fast dephasing times of spins in the solid state~\cite{Fu05a,Dutt06a,Titkov78a}, spin control operating on the picosecond or faster timescale may be necessary.  Such speeds, which are not possible to attain with standard electron spin resonance (ESR) techniques based on microwave sources, can be attained with broadband optical pulses.  One promising ultrafast technique utilizes single broadband pulses detuned from resonance in a three-level $\Lambda$ system~\cite{Clark07a}.  This attractive technique is robust against optical pulse imperfections and does not require a fixed optical reference phase.  Here we demonstrate the principle of coherent manipulation of spins theoretically and experimentally. Using this technique, donor-bound electron spin rotations with single-pulse areas exceeding \PI/4 and two-pulses areas exceeding \PI/2 are demonstrated.  We believe the maximum pulse areas attained do not reflect a fundamental limit of the technique and larger pulse areas could be achieved in other material systems.  This technique has applications from basic solid-state ESR spectroscopy to arbitrary single-qubit rotations~\cite{Clark07a, Economou06a} and bang-bang control~\cite{Viola98a} for quantum computation.

Ultrafast optical techniques have previously been used to study semiconductor spins.  Faraday rotation and differential transmission can be used to passively study spin dynamics on the picosecond and femtosecond timescale~\cite{Crooker97a,Ferrio98a}.  Active techniques to date have utilized either the optical Stark effect~\cite{Gupta01a} or Raman transitions through the \emph{resonant} excitation of an optically active state~\cite{Dutt06a, Greilich06a, Wu07a, Shen07a}.  In contrast to previously demonstrated Raman techniques, in this paper we present a new stimulated Raman transition technique which is based on a single pulse \emph{far-detuned} from the optical transition.  By working off-resonance, decoherence due to real population in the excited state is eliminated without the strict requirement of 2\PI~area pulses~\cite{Economou06a}.  We first introduce the theoretical basis for this technique before presenting the experimental results.

\begin{figure}
\includegraphics[width = 4in]{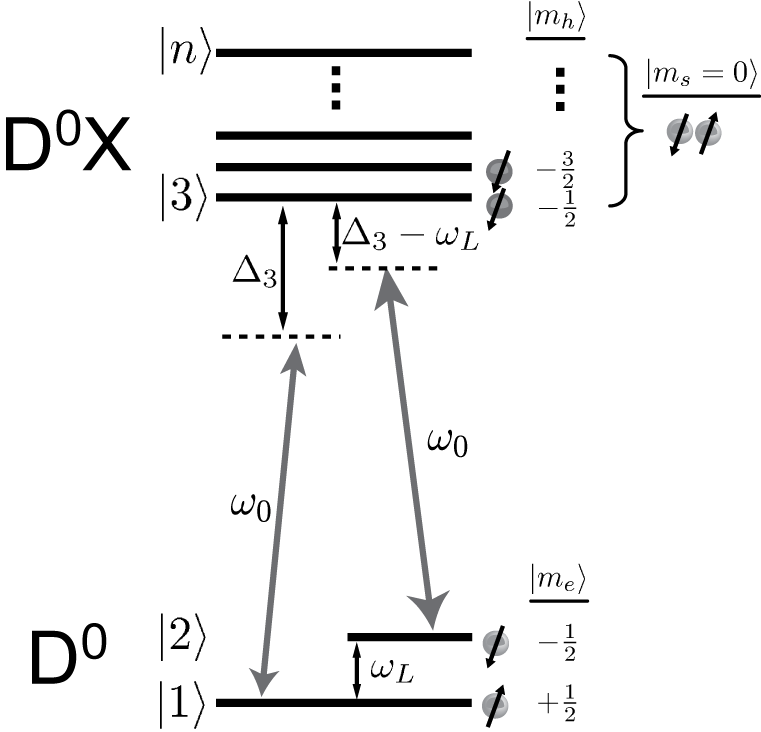}
\caption{$n$-level energy diagram with an applied time-dependent electric field. The two ground levels are split in energy by $\hbar\omega_L$.  An electric field $E$ with energy $\hbar\omega_0$ is applied to the system detuned by the energy $\hbar\Delta_3$.  Energy levels corresponding to the donor bound exciton system are denoted to the right of the diagram and are explained further in the Methods section.}
\label{energylevels}
\end{figure}

 We consider the general $\Lambda$-type system with multiple excited states as depicted in Fig.~\ref{energylevels}.  The lower states are denoted $\ket{1}$ and $\ket{2}$ and in the neutral donor system consist of the spin-up and spin-down states of the bound electron.  These states are coupled via optical dipole transitions to the $(n-2)$ neutral donor-bound exciton states labeled \ket{k}.
To see how the spin can be rotated via a single optical pulse, consider the $n$-level Hamiltonian in the rotating frame
\begin{equation}
H =  \begin{array}{c} |1\rangle \\ |2\rangle \\ |3\rangle \\ \vdots \\ |n\rangle \end{array} \left(\begin{array}{ccccc}0&0&-\frac{\Omega_{31}(t)}{2}&\cdots&-\frac{\Omega_{n1}(t)}{2}\\
0&\omega_L&-\frac{\Omega_{32}(t)}{2}&\cdots&-\frac{\Omega_{n2}(t)}{2}\\
-\frac{\Omega_{31}^*(t)}{2}&-\frac{\Omega_{32}^*(t)}{2}&\Delta_3&\cdots&0\\
\vdots&\vdots&\vdots&\ddots&\vdots\\
-\frac{\Omega_{n1}^*(t)}{2}&-\frac{\Omega_{n2}^*(t)}{2}&0&\cdots&\Delta_n\end{array}\right)
,
\label{eq:nlevel}
\end{equation}
in which $\omega_L$ is the Zeeman splitting of the lower states and $\Delta_k$ is the detuning of the applied pulse from the $\ket{1}~\leftrightarrow~\ket{k}$ transition.  The Rabi frequency $\Omega_{k1}$($\Omega_{k2}$) is the product of the dipole matrix element for the $\ket{1}\leftrightarrow \ket{k} (\ket{2}\leftrightarrow \ket{k})$ transition and the time-dependent electric field amplitude $E$.

This many-level system can be approximated as a two-level spin system by the adiabatic elimination of the upper states, which is valid when the detunings $\Delta_k$ are much larger than other rates in the system~\cite{Harris94a}.  The effective two-level Hamiltonian is given by
\begin{equation}
 H_2
=-\left(
 \begin{array}{cc}
 \frac{|\Omega_1(t)|}{2} & \frac{\Omega_{\text{eff}}(t)}{2} \\
 \frac{\Omega_{\text{eff}}(t)^*}{2} & \frac{|\Omega_2(t)|}{2} - \omega_L
\end{array}\right),
\end{equation}
in which we have defined
\begin{equation}
|\Omega_{1,2}| = \frac{1}{2}\sum_{k>2}^n\frac{|\Omega_{k1,2}|^2}{\Delta_k}.
\label{omega12}
\end{equation}
and an effective Rabi frequency
\begin{equation}
\Omega_{\text{eff}} = \frac{1}{2}\sum_{k>2}^n\frac{\Omega_{k1}\Omega_{k2}^*}{\Delta_k}.
\end{equation}
Population is thus coherently transferred from one lower state to the other at the effective Rabi frequency $\Omega_{\text{eff}}(t)$. Note that the optical phase of the pulse is no longer present in this two-level reduction, eliminating the requirement of a fixed optical reference phase as a clock signal.  If
$\Omega_{\text{eff}}(t) \gg \omega_L$ and $|\Omega_1| = |\Omega_2|$,
the rotation axis will be perpendicular to the magnetic field, and full \PI~rotations with a single pulse are possible.  This condition may often be met by controlling the polarization of the pulse due to the selection rules for the $\ket{1}\leftrightarrow\ket{k}$ and $\ket{2}\leftrightarrow\ket{k}$ transitions.  In material systems in which perpendicular rotations are not possible, large area rotations can still be achieved using multiple pulses by controlling the pulse arrival times over multiple Larmor periods.  For a single-pulse rotation, the phase of the rotation is determined by the phase difference between frequency components separated by the Zeeman frequency within the pulse spectrum and thus is determined by the pulse arrival time~\cite{Clark07a}.  We present the simplified two-level approximation as an intuitive description of the Raman-rotation technique. However, as will be seen below, in a more realistic three-level density matrix model which includes excited state relaxation, high fidelity rotations can still be obtained in a non-adiabatic regime.

One such $\Lambda$-system with multiple excited states that is found in all semiconductors is the neutral donor-bound exciton system.  For the experimental demonstration of the Raman technique, we focused on an ensemble measurement of electrons bound to donors in bulk GaAs.  At liquid helium temperatures the donor electron is bound to the donor impurity creating a neutral donor (\Do).  The~\Do~complex is an attractive potential for excitons (electron-hole pairs) and  the resulting neutral donor-bound exciton (\Dox) consists of the impurity atom, two bound electrons in a spin-singlet state, and a bound hole. The two \Do~spin states and multiple \Dox~states are connected by strong, optical transitions~\cite{Finkman86a} and form the lower and excited states of our $n$-level $\Lambda$-type system as shown in Fig.~\ref{energylevels}.

\begin{figure}
\includegraphics[width = 6in]{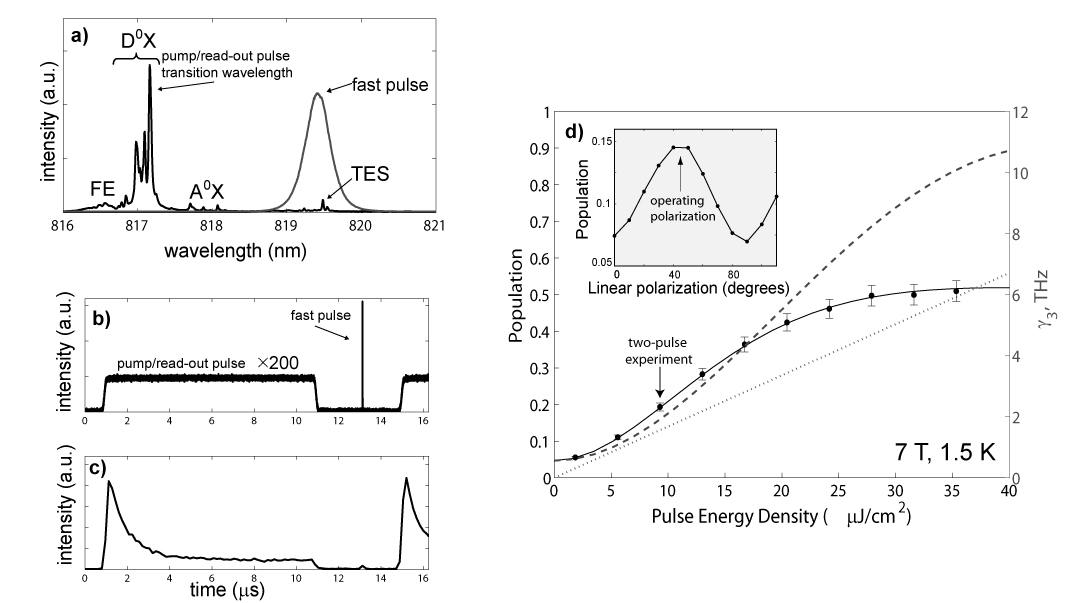}
\caption{Description of the single-pulse experiment and results. {\bfseries a)}~\emph{Black line:} GaAs photoluminescence spectrum with above-band excitation at 815~nm. \emph{Grey line:} Fast pulse spectrum tuned 1~THz from the lowest \Dox~transition. {\bfseries b)} A typical laser-pulse sequence recorded on a GHz photodiode.  {\bfseries c)} A typical PL trace collected from the $\ket{3}\rightarrow\ket{1}$~transition as a function of time. {\bfseries d)} \emph{Markers}:  Experimental population in state \ket{2}~as a function of pulse energy density.   Data were fit to an exponential to determine the initial PL intensity and thus the initial population in state \ket{2}. Error bars are derived from the standard deviation of the fitting coefficients.  A saturation at high energies is observed. The arrow marks the pulse energy used for the two-pulse experiment shown in Fig~\ref{doublepulse}. \emph{Dashed line}: Three-level simulation assuming a constant level-\ket{3}~dephasing rate. \emph{Solid line}: Fit of the data to a three-level simulation using the same parameters as before, except for a state \ket{3} dephasing ($\gamma_3$) with a linear energy dependence. \emph{Dotted line}: Peak value of $\gamma_3$ used for each pulse energy density in the energy dependent dephasing model. \emph{Inset}:~Population measured in state \ket{2} as a function of pulse polarization.}
\label{satcurve}
\end{figure}

In the first experiment we demonstrate population transfer between states \ket{1} and \ket{2} with a single pulse in a 7~T magnetic field.  The experiment consisted of three steps: initialization of the spin population into state \ket{1}, fast-pulse spin transfer, and population readout of state \ket{2}. To initialize the spin state, a continuous-wave field was applied on resonance with the $\ket{2}\leftrightarrow\ket{3}$~transition for 10~$\mu$s.  This transition is the brightest transition in the \Dox~spectrum shown in Fig.~\ref{satcurve}a. At the end of the optical pumping pulse, the state \ket{1}~population was 0.94.  After a 2~$\mu$s delay, which was short compared to the longitudinal relaxation between the lower two states~\cite{Fu06a}, a 2~ps pulse was applied. A typical pulse sequence is shown in Fig.~\ref{satcurve}b.  The pulse was detuned 1~THz below the lowest \Dox~transition. After a second 2~$\mu$s delay the optical pumping pulse was again applied and the population in state \ket{2}~was measured by monitoring the photoluminescence (PL) emitted from the $\ket{3}\rightarrow\ket{1}$~transition at the beginning of the pulse.  The photoluminescence trace as a function of time for a typical case is given in Fig.~\ref{satcurve}c.  The conversion from PL intensity to state $\ket{2}$~population was made by measuring the PL intensity after the system was allowed to return to thermal equilibrium.

As shown above, the relative magnitude of $\Omega_1$ and $\Omega_2$ in Eq.~\ref{omega12}, which can be controlled by the pulse polarization, determines the rotation axis and must be equal for rotations about an axis perpendicular to the magnetic field.  In order to obtain the most efficient population transfer possible, the experiment was first performed at a constant fast-pulse energy for varying pulse polarizations.  The polarization dependence is given in the Fig.~\ref{satcurve}d inset.  Subsequent measurements were performed at the peak of this polarization curve.  In a system in which Rabi oscillations are observed, the amplitude of these oscillations could be used to determine the final rotation axis angle.

In the single-pulse experiment, population transfer was measured as a function of pulse energy by varying the average power of the pulse train.  The results are shown in Fig.~\ref{satcurve}d.  At low pulse energies there is a non-linear increase in population with energy which is characteristic of coherent Raman population transfer.  At higher energies, however, population transfer saturates at a value of 0.5.  This saturation was not expected based on a three-level simulation which includes the experimentally reported relaxation rates for the \Do-\Dox~system (see the Methods section).  In this simulation, a numerical solution of the master equation,
\begin{equation}
\dot{\rho} = -i[H,\rho]+\mathcal{L}(\rho)
\label{eq:master}
\end{equation}
is obtained, where $\rho$ is the three-level density matrix, $H$ is the Hamiltonian given in Eq.~\ref{eq:3level}, and $\mathcal{L}(\rho)$ is the relaxation super-operator given in Eq.~\ref{eq:relax}. In the full density matrix formalism we find that although the evolution is not adiabatic and virtual excitation of the excited state occurs, high fidelity rotations are still possible.  Applying a 2~ps full-width-half-maximum hyperbolic secant pulse with a detuning $\Delta_3$=1~THz and using the definition of fidelity $F = \bra{\psi}\rho \ket{\psi}$, where $\ket{\psi}$ is the desired quantum state, we found that $\pi$ rotations should have been possible with a 0.97 fidelity and $\pi/2$ rotations should have been possible with $>$ 0.99 fidelity in the case that $|\Omega_1| = |\Omega_2|$.  However, a good fit to the experimental data can be obtained using a three-level model which includes a dephasing rate of the excited state ($\gamma_3$) that is linearly dependent on the pulse energy. Although Rabi oscillations were not observed in this first experiment, the non-linear increase in population transfer suggests that small coherent rotations are possible at low powers.

\begin{figure}
\includegraphics[width = 4.5in]{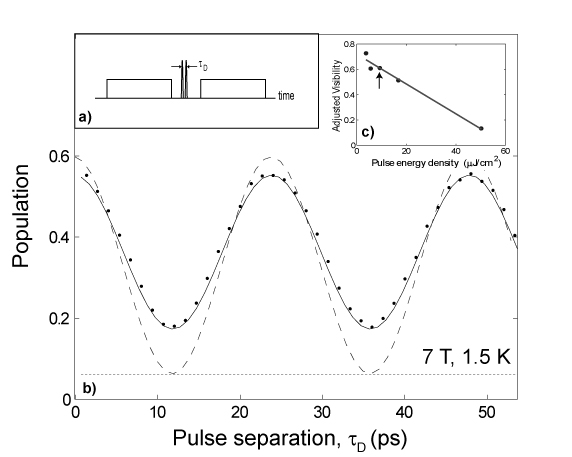}
\caption{Description of the double-pulse experiment and results. {\bfseries a)}~Schematic diagram of the double-pulse experiment pulse sequence.  {\bfseries b)} \emph{Markers}: Population in state \ket{2} after the second pulse as a function of $\tau_D$.  \emph{Solid black line}: The simulation result that uses a level \ket{3}~dephasing of 1.6~THz consistent with the energy dependent dephasing model.  \emph{Dashed line}: Three-level simulation with a 10~GHz level \ket{3}~dephasing rate. \emph{Dotted line}: Residual population remaining in state \ket{2} after the optical pumping pulse.  {\bfseries c)} Pulse energy dependence of the two-pulse visibility.  The solid line is a linear fit.  The visibility is defined as $\frac{I_{max}-I_{min}}{I_{max}+I_{min}}$ in which $I_{max}$ ($I_{min}$) is the maximum (minimum) of the two-pulse visibility curve after subtracting the intensity observed after optical pumping.}
\label{doublepulse}
\end{figure}

In order to measure the coherence of the small angle rotations, a second experiment with two fast pulses was performed. In the double-pulse experiment the single pulse was split into two pulses with a variable delay $\tau_D$ (Fig.~\ref{doublepulse}a).  As the delay between the two pulses increases, the final population in state \ket{2}~oscillates with a period equal to the Larmor period $\tau_L$.  In Fig.~\ref{doublepulse}b we plot the population in state \ket{2}~after two pulses of energy 10~$\mu$J/cm$^2$~were applied, as a function of $\tau_D$.  An oscillation at the Larmor frequency is clearly observed verifying coherent population transfer as well as rotation axis control.  The observed 42~GHz oscillation at 7~T corresponds to a \Do~electron g-factor of $g_e = -0.42$ which is consistent with previous measurements~\cite{Karasyuk94a}.  In contrast to the ideal case, the population in state \ket{2}~never reaches the optically pumped value and perfect destructive interference does not occur.  The finite population left in state \ket{2}~indicates that in addition to coherent population transfer, incoherent population transfer occurs.  The simultaneous fit of the single (Fig.~\ref{satcurve}d) and double-pulse data (Fig.~\ref{doublepulse}b) to the energy-dependent dephasing model indicates a single-pulse rotation of 0.9 radians and a double-pulse rotation of 1.8 radians with fidelities 0.85 and 0.78 respectively.   The two-pulse experiment was performed at several powers with a linear decrease in the visibility observed with increasing power as shown in Fig.~\ref{doublepulse}c.

The saturation of the population transfer in the first experiment and the absence of perfect destructive interference in the second experiment indicate that some fast dephasing mechanism is occurring in the \Do-\Dox~system.  This behavior is not expected from the known parameters in the \Dox~system in the low excitation limit~\cite{Fu05a}. However the single impurity three-level model may not be sufficient to describe the experimental many-exciton system.   We note that in addition to exciting an appreciable virtual \Dox~population during the applied pulse, virtual free excitons are also excited.  As shown in Fig.~\ref{satcurve}a, the \Dox~transitions lie right on the tail of the free exciton transitions.  Unlike the \Dox~transitions, the free-exciton transition is quite broad (THz) and thus additional real excitation due to the picosecond pulse is likely.  We expect the incoherent free exciton excitation to be linear in pulse-energy until the free-exciton transition has saturated.  Once free-excitons are excited, exciton-exciton interactions and exciton-electron interactions~\cite{Piermarocchi02a, Paget81a} could be the source of the observed fast dephasing and a many particle model may be necessary to explain the experiment result.  Dephasing due to multi-exciton interactions should be less of a factor in deeper impurity systems~\cite{Strauf02a} or  charged III-V quantum dot systems~\cite{Dutt06a, Greilich06a, Wu07a} which have much larger exciton binding energies than the GaAs \Dox~system.  In these systems it may be possible to obtain single-pulse large area spin-rotations which would be a valuable tool for spin-based quantum information processing.

\begin{figure}
\includegraphics[width = 3.375in]{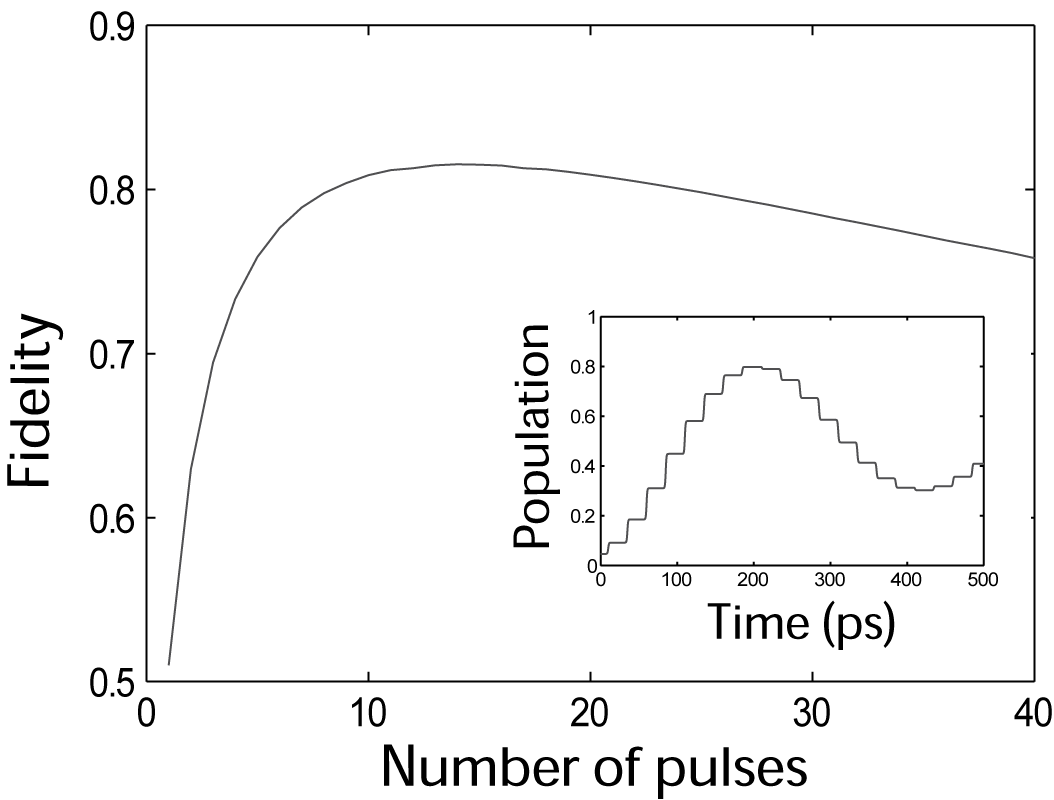}
\caption{Theoretical calculation of the fidelity of $\pi$ pulse vs. number of pulses applied in phase.  \emph{Inset}:  Population in state \ket{2} vs. time as 2~ps pulses are applied in phase.}
\label{cascade}
\end{figure}

Even with the modest population transfer possible in our bulk experimental system, simulations indicate that $\pi$-pulses are still possible if several low energy pulses are applied to the system in phase.  Again, using the three-level model that assumes a level $\ket{3}$ dephasing rate ($\gamma_3$) with a linear power dependence, we find that a $\pi$-rotation is possible with 0.80 fidelity in as few as 8 pulses (200 ps) with energy densities of 5 $\mu$J/cm$^2$~each.  As visible in the Fig.~\ref{cascade} inset, applying small-angle pulses in succession results in a discrete Rabi oscillation curve.  Fig.~\ref{cascade} shows the fidelity of a $\pi$ rotation as a function of the number of pulses applied in phase.  One might expect the fidelity to increase to 1 as more and more weaker pulses are used to perform the rotation.  However the finite $T_2$ decoherence time of 1 ns limits the total number of pulses that can be used.  1~ns is the experimental inhomogeneous dephasing time $T_2^*$ in the system~\cite{Fu05a} and the decoherence time could be much longer for a single spin system.  While the multiple-pulse technique is significantly slower than the single-pulse technique, it may prove valuable in systems in which high-area rotations are desired yet low powers are necessary.  Such large area rotations are necessary for single gates in quantum computation and can also aid in the suppression of decoherence~\cite{Viola98a}.  However, for general ESR techniques, such as spin-echo,  small area pulses are all that is necessary for determining the fundamental homogeneous decoherence time T$_2$~\cite{Rosatzin90a} in a material.  Thus this Raman fast-pulse technique as experimentally demonstrated can immediately be applied for these studies.

\section{Methods}
\subsection{The \Dox~system}

The sample studied consisted of a 10~$\mu$m GaAs layer with a donor density of $5\times10^{13}~$cm$^{-3}$~on a 4~$\mu$m Al$_{0.3}$Ga$_{0.7}$As layer grown by molecular-beam epitaxy on a
GaAs substrate. The sample was mounted strain-free in a magnetic cryostat in a liquid helium bath.  The magnetic field was parallel to the $<110>$ crystallographic axis.  The magnetic field was perpendicular to the excitation and collection paths. The signal was collected from a 20~$\mu$m spot with an estimated $10^5$ donors contributing to the signal.

In the applied magnetic field, the neutral donor electron splits into two levels denoted by the magnetic quantum number $m_e = \pm\frac12$ as shown in Fig.~\ref{energylevels}.  The bound electron g-factor $g_e$ is -0.40 to -0.46 depending on the strength of the magnetic field and the crystal orientation with respect to the magnetic field~\cite{Karasyuk94a}.  The excited state \Dox~energy level structure is much more complex and although it is not fully understood a detailed study has been performed by Karasyuk et al.\cite{Karasyuk94a}.   The two electrons in the complex form a spin singlet denoted in Fig.~\ref{energylevels} by $|m_s = 0 \rangle$.  The energy of the ~\Dox~state is thus determined by the spin of the bound hole $m_h = \pm\frac12, \pm\frac32$ as well as the hole's effective mass orbital angular momentum $L$.  In our system we have identified the ground and first excited state of the \Dox~complex as the $|L=1, m_h = -\frac12\rangle$ and the $|L = 0, m_h = -\frac32\rangle$ states respectively.

A GaAs photoluminescence spectrum with above-band excitation at 815~nm can be seen in Fig.~\ref{satcurve}a (blue curve).  Only the horizontal polarization is collected to resolve more of the \Dox~transitions and \Dox~linewidths are instrument resolution limited.  Also observed in the spectrum are the free exciton transitions, the acceptor bound-exciton (\Aox) transitions, and the \Dox~two-electron satellite transitions (TES).  TES transitions occur when the \Dox~relaxes into a \Do~excited state.

High resolution photoluminescence excitation (PLE) spectroscopy as well as the location of the TES lines indicate that the primary donor in our sample is silicon with a much weaker ($<10\times$) concentration of sulfur.  PLE spectroscopy on the sample shows that the inhomogeneous linewidth is $<10$~GHz on the broadest lines~\cite{Fu05a}.  This linewidth is much narrower than the 1~THz detuning and thus should not affect the fidelity of the pulse rotations.

\subsection{More on the experimental pulse sequence}
A typical fast-pulse sequence detected on a GHz photodiode is shown in Fig.~\ref{satcurve}b in which the pump/read-out pulse intensity is 200 times the intensity used in the experiment.  The optical pumping laser was a Coherent Ti:Sapphire 899-29 continuous-wave (CW) laser modulated by an AOM and was polarized parallel to the magnetic field.  Depending on the experiment the pump intensity ranged from 5 to 15~$\mathrm{\mu W}$ and the spot size was approximately 120~$\mathrm{\mu m}$.  The fast pulse was provided by a picosecond mode-locked laser with a repetition rate of 80~MHz.  Single pulses were picked by an EOM every 10 to 15 ~$\mathrm{\mu s}$ depending on the particular experiment.  The EOM extinction ratio ranged from 60-100 and an additional 2~$\mathrm{\mu s}$ extinction envelope was provided by an AOM (extinction ratio of 1000).  The fast-pulse polarization was 45 degrees from the magnetic field axis.  The fast-pulse laser spot size was 80~microns.

\subsection{Three-level theoretical model}

The data in Figs.~\ref{satcurve}d and ~\ref{doublepulse}b are fit to a three level density matrix model which solves the master equation given in Eq.~\ref{eq:master}.  The three-level Hamiltonian in the rotating frame is given by
\begin{equation}
H_3 = \left(\begin{array}{ccc}0&0&-\frac{\Omega_{31}(t)}{2}\\
0&\omega_L&-\frac{\Omega_{32}(t)}{2}\\
-\frac{\Omega_{31}^*(t)}{2}&-\frac{\Omega_{32}^*(t)}{2}&\Delta\end{array}\right),
\label{eq:3level}
\end{equation}
and the relaxation operator $\mathcal{L}(\rho)$ is given by \begin{widetext}
\begin{equation}
 \mathcal{L}(\rho) = \left( \begin{matrix}
 -\Gamma_{12}\rho_{11}+\Gamma_{21}\rho_{22} +
 \Gamma_{3}\rho_{33}  &

-(\frac{\Gamma_{12} + \Gamma_{21}}{2} +  \gamma_{2}) \rho_{12} &

-(\frac{\Gamma_{12} + 2\Gamma_{3}}{2} + \gamma_{3}(t))\rho_{13} \\

-(\frac{\Gamma_{12} + \Gamma_{21}}{2} +  \gamma_{2}) \rho_{21} &

\Gamma_{12}\rho_{11} - \Gamma_{21}\rho_{22} + \Gamma_{3}\rho_{33}
&

-(\frac{\Gamma_{21} + 2\Gamma_{3}}{2} + \gamma_{3}(t))\rho_{23} \\

-(\frac{\Gamma_{12} + 2\Gamma_{3}}{2} +
\gamma_{3}(t))\rho_{31} &

-(\frac{\Gamma_{21} + 2\Gamma_{3}}{2} +
\gamma_{3}(t))\rho_{32} &

-(2\Gamma_{3})\rho_{33}

\end{matrix} \right)
\label{eq:relax}
\end{equation}
\end{widetext}
 in which $\Gamma_{12}$ ($\Gamma_{21} =
\Gamma_{12}e^\frac{E_{12}}{kT}$) is the longitudinal relaxation
rate from $|1\rangle\rightarrow|2\rangle$
($|2\rangle\rightarrow|1\rangle$), $2\Gamma_{3}$
is the total radiative relaxation from $|3\rangle$, $\gamma_2$ is the transverse relaxation rate between $|1\rangle$ and $|2\rangle$, and
$\gamma_{3}(t)$ is the level $|3\rangle$ dephasing.  For electrons bound to neutral donors in GaAs, the relevant relaxation parameters are the excited state radiative relaxation rate $2\Gamma_{3}=(1$~ns$)^{-1}$~\cite{Finkman86a}, the lower state decoherence time $T_2=1$~ns~\cite{Fu05a}, and a Zeeman splitting of $\omega_L=42$~GHz in a 7~T field.  In the constant level-\ket{3}~dephasing model $\gamma_3(t)=10$~GHz~\cite{Karasyuk94a, Fu05a}.  Three level simulations show that even in this non-adiabatic regime, \Dox~ population is only virtually excited during the pulse duration.  Pulse fidelities are high as long as all dephasing mechanisms in the system are slow compared to the pulse time. In the intensity dependent model the peak value of $\gamma_3(t)$ is given in Fig.~\ref{satcurve}d. The model fits the data if pulse energies are a factor of 0.8 less than the energies in the experiment. This discrepancy could be explained by our incomplete knowledge of the \Dox~ dipole matrix elements. The model includes only one excited state with a relaxation rate equal to the experimental total relaxation rate for the many level \Dox~state at zero magnetic field.  In reality our fast pulse is interacting with many levels in a 7~T field.

\section{Author Contributions}
C.R.S. and M.C.H. provide the sample.  K-M.C.F and S.M.C. designed, built, and performed the experiment.  K-M.C.F. analyzed and modeled the experimental data and did the majority of the writing with significant contributions from S.M.C. S.M.C. contributed the theoretical multiple small-angle pulse model. C.S. and Y.Y. contributed throughout the experimental process from the conception to the writing through useful suggestions, comments, and discussions.

Correspondence should be addressed to K-M.C.F. at kai-mei.fu@hp.com.

\section{Acknowledgements}

The authors thank Steve Harris and Jason Pelc for helpful discussions.  S.M.C. was partially supported by the HP Fellowship Program through CIS.  This work was financially supported by the MURI Center for photonic quantum information systems (ARO/ARDA Program DAAD 19-03-1-0199), JST/SORST program for the research of quantum information systems for which light is used, and the University of Tokyo (CINQIE) Special Coordination Funds for Promoting Science and Technology, the ``Qubus quantum computer program'' MEXT and NICT program on Quantum Repeaters.

\bibliographystyle{nature}

\newpage

\end{document}